\documentclass[twocolumn,longbib]{aastex701}

\usepackage{graphicx}
\usepackage{bm}
\usepackage{amsmath}

\newcommand{\rem}[1]{ } 

\begin{document}


\title{Theory of striped dynamic spectra of the Crab pulsar high-frequency interpulse}

\author[0000-0001-5987-2856]{Mikhail V. Medvedev}
\affiliation{Department of Physics and Astronomy, University of Kansas, Lawrence, KS 66045}
\affiliation{Laboratory for Nuclear Science, Massachusetts Institute of Technology, Cambridge, MA 02139}
\email[show]{medvedev@ku.edu}



\begin{abstract}
A theory of the spectral ``zebra" pattern of the Crab pulsar's high-frequency interpulse (HFIP) radio emission is developed. The observed emission bands are interference maxima caused by multiple ray propagation through the pulsar magnetosphere. The high-contrast interference pattern is the combined effect of gravitational lensing and plasma de-lensing of light rays. The model enables space-resolved tomography of the pulsar magnetosphere, yielding a radial plasma density profile of $n_{e}\propto r^{-3}$, which agrees with theoretical insights. We predict the zebra pattern trend to change at a higher frequency when the ray separation becomes smaller than the pulsar size. This frequency is predicted to be in the range between 42 GHz and 650 GHz, which is within the reach of existing facilities like ALMA and SMA. These observations hold significant importance and would contribute to our understanding of the magnetosphere. Furthermore, they offer the potential to investigate gravity in the strong field regime near the star's surface.
\end{abstract}




\section{Introduction}

Pulsars are highly magnetized neutron stars that emit pulsed electromagnetic radiation. The properties and theories of pulsars are comprehensively reviewed by \citet{PK22}. Among the over 3,700 known pulsars, the Crab pulsar stands out as one of the brightest and most extensively studied. It emits two pulses per rotation period that are observed across the entire electromagnetic spectrum, from radio waves to X-rays. The temporal coincidence of the pulses at different wavelengths suggests that the observed emissions originate from the same physical location. It is now widely accepted that the Crab pulsar emissions originate from within the current sheet, outside the light cylinder \citep{BS10}. The mechanism underlying this process involves magnetic reconnection and subsequent violent interactions of plasmoids within the current sheet \citep{P+19}. This model implies that an individual pulse emission, including giant flares, is composed of a large number of (blended in time) ``nanoshots" — bright sub-pulses of a few nanosecond duration each — thereby explaining their high temporal variability and broadband spectra.

In this paper, we are interested in the radio emission of the object in the GHz frequency range. The observed broadband spectrum is a general feature of the main pulse, low-frequency interpulse, and several emission components in the frequency range between approximately 1 GHz and 30 GHz, as studied in Refs. \citep{HE07, HE+16a, HE+16b}. In contrast, the spectral pattern of the high-frequency interpulse (HFIP), observed between approximately $\nu \sim 5$ GHz and $\nu \sim 30$ GHz, is remarkably different and represents a sequence of emission bands resembling the ``zebra" pattern. This peculiar spectral pattern was first reported in 2007 and subsequently studied in great detail \citep{HE07, HE+16b}. 

The following are the known properties of HFIP emission that any successful model should account for:
\begin{itemize}
\item The presence of peculiar spectral features in the form of regular emission bands.
\item The proportional separation of the bands and the ``6\% rule". The frequency difference between the nearest bands is proportional to the band frequency, resulting in $\Delta\nu \approx 0.057\nu$.
\item The persistence of the pattern. There has not been observed a single HFIP without spectral bands.
\item The stability of the pattern. The band positions can remain stable for as long as a day, although they can also vary from one pulse to the next.
\item The HFIP is nearly 100\% linearly polarized, and the position angle is stable and does not change over many pulses.
\item The HFIP has a variable and often larger dispersion measure (DM) compared to the main pulse.
\item No Faraday rotation within the system has been reported. 
\end{itemize}

Despite substantial theoretical efforts over the subsequent fifteen years, no satisfactory mechanism has been proposed to elucidate the HFIP puzzle. The proposed models either involve emission mechanisms generating multiple harmonics (e.g., cyclotron or maser emissions) or invoke propagation effects (e.g., interference at a source, within a current sheet, or wave modulation instability). The former class of models predicts an incorrect uniform band spacing. The latter class necessitates extremely high wave phase coherence and source stability, which are unrealistic to anticipate in a highly turbulent pulsar wind medium. A comprehensive review and details on recent models can be found in Refs. \citep{HE16c, Benachek+24, S+24}.

Recently, \citet{M24} proposed an attractive model that explains the ``zebra" band structure as the spectral interference pattern resulting from multiple light propagation paths within the pulsar magnetosphere. In this paper, we further elaborate upon the model to incorporate the influence of general relativity, which appears to be of significant importance. Here, we calculate the light ray paths from the first principles and demonstrate that the pulsar system is equivalent to the Young's double-slit setup, which produces the high-contrast interference fringe pattern with visibility of order unity, in contrast to diffraction. Furthermore, the deduced plasma density distribution within the magnetosphere exhibits an inverse-cube-law, which is anticipated for a dipolar magnetic field. 

The remainder of the paper is organized as follows. Section \ref{s-model} presents the details of the model. In Section \ref{s-prop}, the light ray paths through the plasma in a Schwarzschild geometry are computed. In Section \ref{s-int}, the interference of these rays is studied. Section \ref{s-disc} presents important discussion and predictions. Section \ref{s-concl} summarizes major results of the paper.

\section{Model}
\label{s-model}

\begin{figure}
\vspace{15pt}
\center
\includegraphics[scale=0.25]{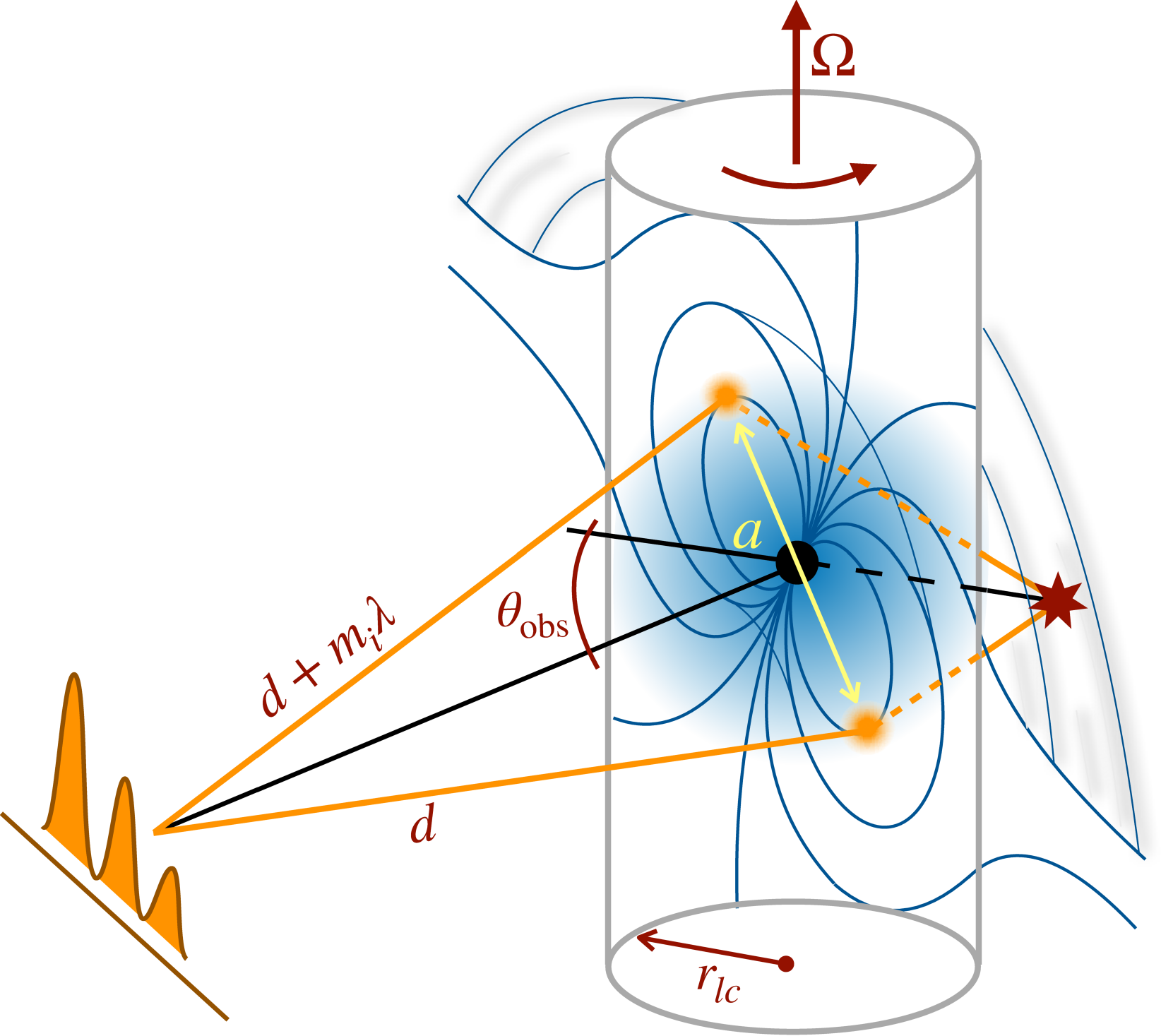}
\caption{Overall geometry of the system. }
\label{f-pulsar}
\end{figure}

The primary assumption of the model is that the broadband radio emission source of HFIP is located behind the pulsar, as illustrated in Fig. \ref{f-pulsar}. The precise origin of the source remains uncertain. Nevertheless, it is reasonable to assume that the mechanism is analogous to that observed in the main pulse and the low-frequency interpulse \citep{P+19}. In principle, although less likely, this could be a reflected or refracted radio emission originating from a distinct location, such as a polar cap. 

Electromagnetic waves, propagating in magnetized plasma quasi-perpendicular to the magnetic field have two distinct polarizations, the ``ordinary modes'' and ``extraordinary modes'' (O-modes and X-modes, for short). These two polarizations possess different propagation characteristics. 

The X-mode possesses a component of the wave electric field perpendicular to the magnetic field. Consequently, it can resonate with gyrating electrons and positrons, resulting in significant absorption. It is widely recognized that cyclotron resonant absorption severely restricts wave propagation through a magnetosphere \citep{BS76, M+82}. Absorption typically occurs near the light cylinder, where the wave frequency and the cyclotron frequency coincide. Although the resonance region is typically confined to a small volume, the absorption cross-section is very high, resulting in a substantial overall optical depth.
The optical depth in the Crab magnetosphere is approximately $\tau_{\rm abs} \sim 0.1{\cal M}(\nu/{\rm 30~GHz})^{-1/3}$ \citep[cf. Eq. (8) in Ref.][]{RG05}. Consequently, the X-mode is strongly absorbed ($\tau_{\rm abs} \gtrsim 1$) in a plasma with multiplicity ${\cal M} \gtrsim 10$, irrespective of whether the propagation is perpendicular or oblique. In the Crab pulsar environment, ${\cal M} \sim 10^4 - 10^5$ on open field lines and possibly lower on closed field lines \citep{PK22}. 

It is crucial to emphasize that the nearly 100\% polarization of HFIP is naturally explained by the strong absorption of one of the two modes. The constancy of the position angle can be attributed to pure geometry: the radio source is consistently located behind the magnetosphere at the same rotation phase. Variations in the position angle and circular polarization, if any, can be affected as plasma normal modes propagate through the dynamical magnetospheric plasma \citep{PL00, BP12}, as well as caused by variability of the source.

In contrast, the O-mode propagating perpendicular to the magnetic field has its wave electric field aligned with the ambient magnetic field. Consequently, it perceives plasma but not the magnetic field. Consequently, it does not undergo cyclotron absorption and cannot propagate at a frequency below the plasma frequency (it is said to have a cutoff at the plasma frequency). The plasma index of refraction for the perpendicular-propagating O-mode is identical to that of the unmagnetized plasma:
\begin{equation}
n_{\rm pl}^{2}(r,\omega)\equiv\left(\frac{kc}{\omega}\right)^2=1-\frac{\omega_p^2}{\omega^2}=1-\frac{4\pi e^2 n_e(r)/m_e}{\omega^2},
\label{n2pl}
\end{equation}
where $\omega$ and $k$ are the frequency and wave number of the wave, $c$ is the speed of light, $\omega_p$ is called the nonrelativistic plasma frequency, $e$ is the elementary charge, and $n_{e}$ is the lepton (electron and positron, if present) number density. In relativistic electron-positron plasmas, one should replace the electron mass $m_{e}$ with the ensemble-averaged quantity $m_{e}\langle\gamma_{e}^{-3}\rangle^{-1}$ with $\gamma_{e}$ being the particles' Lorenz factor \citep{Gedalin+98}. This factor can be heuristically understood as the effect accounting for the relativistic parallel inertia. This effect arises when the acceleration due to the transverse-propagating ordinary-wave electric field is parallel (or antiparallel) to the particle velocity, which in turn is along the magnetic field. Plasma on the open field lines is highly relativistic, $\langle\gamma_{e}^{-3}\rangle^{-1}\gg1$, as it has a large bulk Lorentz factor and also can have relativistically large thermal spread at large distances.
On the closed field lines, however, the plasma is nonrelativistic because of cooling and magnetic confinement, $\gamma_{e}\sim1$. Consequently, we can disregard this factor henceforth. It can always be accounted for later, if necessary.

It is crucial to emphasize that we assume the propagation of O-mode waves is strictly perpendicular to the magnetospheric field. This assumption holds true for the waves propagating in the plane of the magnetic equator of a magnetic dipole, which serves as a model for the magnetic field of a neutron star. Consequently, this assumption imposes a constraint on the location of the source, yet it remains fully consistent with the source being situated at the current sheet. Furthermore, propagation in the equatorial plane inherently leads to the disappearance of Faraday rotation, irrespective of the plasma's composition.
(Note: Generally, Faraday rotation exists even in electron-positron plasma, provided it is non-neutral. The charge imbalance induced by the spinning magnetosphere is equal to the Goldreich-Julian density.)

\section{Propagation}
\label{s-prop}

In this section, we employ the natural system of units, $\hbar=c=G=1$. Propagation of light through plasma in a strong gravitational field has been studied extensively, see for example Refs. \citep{dF71, BB03a, BB03b, TB-K13, AAA15, PT17, BB24}. Here we present relevant theoretical calculations of the ray trajectories, which will be used in the next section. 

The plasma dispersion of the O-mode, $\omega^{2}=\omega_p^2+k^{2}$, in the covariant form reads
\begin{equation}
g^{\mu\nu}p_\mu p_\nu+\omega_p^2=0,
\label{disp}
\end{equation}
where the zeroth component of the 4-momentum is $p_{t}=-\omega$ and the spatial components are $p_{i}=k_{i},\ i=1,2,3$. Here $\omega=\omega({\bf x})$ and $k=k({\bf x})$ are the local frequency and wavenumber of the electromagnetic wave, and $g^{\mu\nu}$ is the spacetime metric tensor.  
The dispersion relation induces a Hamiltonian for our system:
\begin{equation}
H({\bf x},{\bf p})=\frac{1}{2}\left[g^{\mu\nu}({\bf x})p_\mu p_\nu+\omega_p^2({\bf x})\right].
\label{H}
\end{equation}
%
Then, geodesic equations for light rays are
\begin{align}
&\frac{d x^\mu}{d \lambda}=\frac{\partial H}{\partial p_\mu},
\label{HamE1}\\
&\frac{d p_\mu}{d \lambda}=-\frac{\partial H}{\partial x^\mu},
\label{HamE2}\\
&H=0,
\label{HamE3}
\end{align}
where $\lambda$ is the affine parameter along the ray trajectory. 

We assume Schwarzschild geometry
\begin{align}
ds^2&=g_{\mu\nu}dx^\mu dx^\nu\\
&=-\left(1-\frac{2m}{r}\right)dt^2+\left(1-\frac{2m}{r}\right)^{-1}dr^2+r^2d\Omega^2.
\end{align}
Here $d\Omega^2=(d\theta^{2} + \sin^{2}\theta\, d\phi^{2})$ is the metric on the two-sphere, $\theta$ and $\phi$ are the azimuthal and polar angles. Thereby, the rotation of the neutron star is neglected. In our units, the gravitational (Schwarzschild) radius is given by $r_g = 2m$, where $m$ is the mass of the neutron star. The Hamiltonian, Eq. \eqref{H}, in Kerr spacetime reads
\begin{align}
H({\bf x},{\bf p})&=\frac{1}{2}\left[\left(1-\frac{2m}{r}\right)p_{r}^{2}
+\frac{p_{\theta}^{2}}{r^{2}}+\frac{p_{\phi}^{2}}{r^{2}\sin^{2}\theta}
\right.\nonumber\\
&\qquad \left.-\left(1-\frac{2m}{r}\right)^{-1}p_{t}^{2}+\omega_{p}^{2}({\bf x})\right].
\end{align}

Let $S$ be action, then $p_{\mu}=\partial S/\partial x^{\mu}$. Consequently, Eq. \eqref{disp} transforms into the Hamilton-Jacobi equation, which can be solved using separation of variables, $S=-E t+L \phi +S_{r}(r)+ S_{\theta}(\theta)$. Here, the two constants of motion represent conserved quantities, specifically energy and angular momentum: 
\begin{align}
&E=-\partial S/\partial t=-p_{t}=\omega_{\infty},\label{E}\\
&L=\partial S/\partial \phi=p_{\phi}=k b= \omega_{\infty}b,\label{L}
\end{align}
where $b$ is the impact parameter, the constant $\omega_{\infty}=\omega(r\to\infty)$ is the frequency of radiation at infinity, and the last equalities in Eqs. \eqref{E}, \eqref{L} hold true because $\omega_p(r)=0$ as $r\to \infty$. Hereafter we assume that $\omega_p(r)$ solely depends on the radial coordinate, thereby maintaining spherical symmetry. 

The third constant of motion that emerges from the separation of variables in the Hamilton-Jacobi equation, referred to as Carter's constant, appears as an identity and is expressed as
\begin{align}
{\cal C}&=p_{\theta}^{2}+L\cot^{2}\theta \label{C} \\
&=-r^{2}\left[\left(1-\frac{2m}{r}\right)p_{r}^{2}+\frac{L^{2}}{r^{2}}
-\frac{E^{2}}{\left(1-\frac{2m}{r}\right)}+\omega_{p}(r)\right]. \nonumber
\end{align}
By employing spherical symmetry, we can assume, for convenience and without loss of generality, that motion starts in the equatorial plane, $\theta=\pi/2$, with vanishing $\theta$-momentum at infinity, $p_{\theta,\infty}=0$. Then Carter's constant vanishes identically. Note that Eq. \eqref{C} with ${\cal C}=0$ readily determines the radial momentum:
\begin{align}
p_{r}^{2}=\left(1-\frac{2m}{r}\right)^{-2}\left[E^{2}-\left(1-\frac{2m}{r}\right)\left(\frac{L^{2}}{r^{2}}+\omega_{p}^{2}(r)\right)\right].
\label{pr}
\end{align}

Substituting Eqs. \eqref{E}--\eqref{pr} into the Hamilton equations, Eqs.\eqref{HamE1}--\eqref{HamE3}, the light ray equations can be written as
\begin{align}
&\frac{d r}{d \lambda}=\sqrt{\omega_\infty^2-\left(1-\frac{2m}{r}\right)\left(\frac{\omega_\infty^2b^2}{r^2}+\omega_p^2\right)},
\label{drdl}\\
&\frac{d \theta}{d \lambda}=0,
\label{dthdl}\\
&\frac{d \phi}{d \lambda}=\frac{\omega_\infty b}{r^{2}},
\label{dphdl}\\
&\frac{d t}{d \lambda}={\omega_\infty}\left/{\left(1-\frac{2m}{r}\right)}\right.
\label{dtdl}.
\end{align}
These equations contain three constants of motion that separate the variables in the Hamilton-Jacobi equation. Two of them, $\omega_{\infty}$ and $b$ (in the combination $\omega_{\infty}b$), represent the conservation of energy and angular momentum, respectively. The third one (Carter's constant), together with Eq. (\ref{dthdl}), yields that motion is restricted to a plane, chosen to be the equatorial plane, $\theta(\lambda)=\theta(0)=\pi/2$, and $p_{\theta}=0$ everywhere, not just at infinity. 

To proceed further, we need to employ a model of the electron-positron plasma density distribution. Motivated by the recent finding suggesting a power-law distribution in the magnetosphere of the Crab pulsar \citep{M24}, we assume $\omega_p^2\propto n_{e}(r)\propto r^{-\kappa}$ with $\kappa$ being a constant to be determined. This yields the following parameterization of the plasma index of refraction, Eq. \eqref{n2pl}:
\begin{equation}
n_{\rm pl}^{2}(r,\omega_{\infty})=1-\left(\frac{r_{0}(\omega_{\infty})}{r}\right)^{\kappa}.
\end{equation}
Here $r_0(\omega_{\infty})$ is the frequency-dependent plasma ``reflection radius'' in the absence of gravity. It is defined as the radius at which the plasma frequency equals the wave frequency at infinity, $\omega_{p}(r_0) = \omega_{\infty}$. Thus, in the absence of gravitational blueshift, this is the point where the wave would be reflected, as indicated by the condition $n_{\rm pl}(r_0, \omega_{\infty}) = 0$. Consequently,
\begin{equation}
r_0^\kappa(\omega_{\infty})=r_*^\kappa\,\frac{\omega_{p*}^2}{\omega_\infty^2},
\label{r0}
\end{equation}
where we chose to normalize quantities by their values at the neutron star surface, namely $r_*$ is the neutron star radius and $\omega_{p*}\equiv\omega_p(r_*)$ is the plasma frequency at the neutron star's surface.

\begin{figure}
\vspace{15pt}
\center
\includegraphics[scale=0.9]{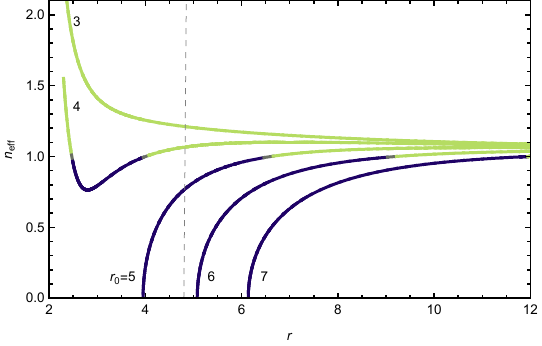}
\caption{The effective index of refraction as a function of distance, for various values of the so-called ``reflection radius'' parameter $r_{0}=3,\dots, 7$ and $\kappa=3$. All distances are in units of mass $m$, so that $r_{g}=2$ and the Crab pulsar radius is $r_{*}\simeq 4.8$ (indicated by the vertical dashed line). Bright colors denote $n_{\rm eff}>1$ and dark colors correspond to $n_{\rm eff}<1$.
}
\label{f-neff}
\end{figure}

The equation that describes the geometric path of a light ray in a medium, here referred to as the ``Crystal Ball equation" (see Appendix for discussion), is straightforwardly obtained from Eqs. (\ref{drdl}) and (\ref{dphdl}): 
\begin{equation}
\frac{d \phi}{d r}=\frac{b}{r\sqrt{1-{2m}/{r}}\sqrt{r^2 n_{\rm eff}^2-b^2}},
\label{dfdr}
\end{equation}
where the effective index of refraction is 
\begin{equation}
n_{\rm eff}=\sqrt{\left(1-\frac{2m}{r}\right)^{-1}-\left(\frac{r_0}{r}\right)^\kappa}.
\end{equation}

Fig. \ref{f-neff} illustrates the behavior of the effective index of refraction, $n_{\rm eff}$, as a function of distance. The figure depicts a range of values for the reflection radius parameter $r_{0}$, which varies between 3 and 7. The figure considers a representative value of $\kappa=3$, which corresponds to the Goldreich-Julian density in the dipolar magnetic field. All distances are measured in units of mass, $m$. Consequently, $r_{g}=2$, and the Crab pulsar radius (for a mass of $M_{*}\sim 1.4 M_{\odot}$ and a radius of $R_{*}\simeq 10$ km) is approximately $r_{*}\simeq4.8$. This value is roughly equivalent to 2.4 Schwarzschild radii. Bright colors of the curves indicate regions where $n_{\rm eff}>1$, signifying the dominance of gravitational lensing. Conversely, dark colors correspond to regions where $n_{\rm eff}<1$, indicating the dominance of defocusing by plasma.

We observe an intriguing interplay between gravity and plasma dispersion, especially at short distances when $r\sim r_{0}\sim r_{g}$. This interplay results in a non-monotonic index of refraction. This phenomenon may hold significance for the study of light propagation near a black hole. Further exploration of this effect is warranted in a separate context. In contrast, for our neutron star system, which has a substantially larger size, the refraction index behavior is relatively straightforward. It remains below unity in the vicinity of $r_{0}$, indicating the predominance of plasma dispersion, and increases above unity at greater distances, where gravitational lensing effect dominates.  

The geometrical path of the light ray, $\phi(r)$, is determined by the straightforward integration of Eq. \eqref{dfdr}. The minimum radius, or the distance of the closest approach, $r_{m}$, is identified by the vanishing denominator. Consequently, it is implicitly determined, as a function of $b$, as a positive root of the equation
\begin{equation}
r_{m}n_{\rm eff}(r_{m})=b.
\label{rm}
\end{equation}
Note that this equation is polynomial for rational values of $\kappa$ and transcendental for irrational values. By symmetry of the trajectory with respect to the $r=r_{m}$ point, the total change in the angle $\phi$ is
\begin{equation}
\phi_{\rm tot}=2\int^{r_{m}}_{\infty}\frac{b\, dr}{r\sqrt{1-{2m}/{r}}\sqrt{r^2 n_{\rm eff}^2-b^2}}.
\label{ftot}
\end{equation}
A straight, non-deflected path has a total angle change $\phi_{\rm tot} = \pi$. Consequently, the deflection angle $\hat\alpha$ is determined as $\hat\alpha \equiv \phi_{\rm tot} - \pi$. Positive values of $\hat\alpha$ indicate focusing, while negative values correspond to defocusing. 

\begin{figure}
\vspace{15pt}
\center
\includegraphics[scale=0.9]{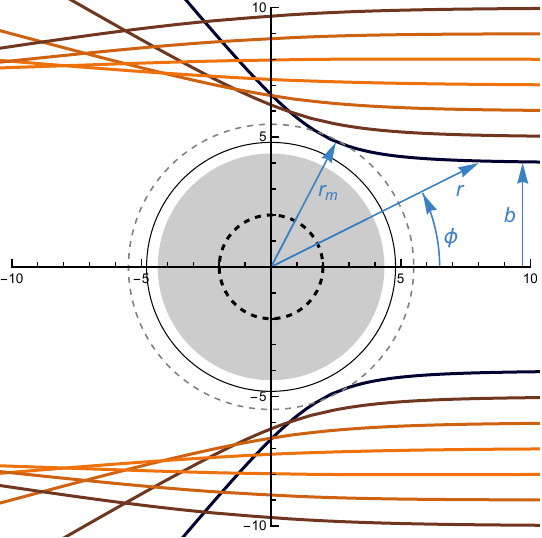}
\caption{Deflection of parallel light rays propagating from right to left for various impact parameters, $b=4,\dots,10$, and for $r_{0}=5.5$ and $\kappa=3$. Brighter colors indicate weak deflection, while darker tones represent strongly deflected rays. The $(r,\phi)$ coordinates of the innermost ray, along with its impact parameter, $b$, and the distance of the closest approach, $r_{m}$, are depicted. Also, plotted are: the Schwarzschild radius, $r_{g}=2$ (smallest dashed circle), the non-transparent region  when gravitational blue-shift of $\omega$ is accounted for (gray disk), the neutron star size, $r_{*}\simeq4.8$ (solid black circle), and the reflection radius, $r_{0}$, defined by Eq. \eqref{r0} (large dashed gray circle). All distances are measured in units of the neutron-star mass $m$. 
}
\label{f-rays}
\end{figure}

Fig. \ref{f-rays} illustrates the deflection of light rays from a light source located at infinity on the right and propagating leftward. The figure depicts a range of values for the impact parameter $b$, which varies between 4 and 10. The light rays are colored by their deflection angle with brighter colors indicating weaker deflections, whereas darker colors indicate stronger deflections. Other parameters are fixed at $r_{0}=5.5$ and $\kappa=3$. For the innermost ray (thick black curve), the coordinate system, $(r,\phi)$, its impact parameter, $b$, and the distance of the closest approach, $r_{m}$, are depicted. Additionally shown with circles are the gravitational radius $r_{g}=2$ (dashed black), the neutron star radius, $r_{*}\simeq4.8$ (solid black), and the reflection radius, $r_{0}$ (dashed gray). We also depict the region of non-transparency (gray disk), which is different from $r_{0}$ (which is defined for $\omega_{\infty}$), because of the gravitational blueshift of the wave frequency as it enters deeper into the gravitational field, $\omega>\omega_{\infty}$, and becomes maximum at $r_{m}$. Consequently, the distinction between the actual reflection radius, which corresponds to the edge of the gray disk, and the reflection radius $r_{0}$, for which gravitational effects are disregarded, elucidates the magnitude and significance of the gravitational blueshift's impact. 

As observed, rays with minimal deflections, $|\hat\alpha| \ll 1$, are present for sufficiently large values of $b$, i.e., at considerable distances. Consequently, the analytical treatment can be further simplified by assuming $r \gg 2m$ and $r \gg r_0$. We evaluate the applicability of the latter assumption in Section \ref{s-disc}. These assumptions correspond to the limit of weak gravitational field and weak plasma refraction. In this limit, referred to as the ``far-away approximation," the Crystal Ball equation simplifies to a classical form: 
\begin{equation}
\frac{d \phi}{d r}=\frac{b}{r\sqrt{r^2 n^2(r)-b^2}}.
\label{cbe}
\end{equation}
This equation describes the trajectory of a light ray's path through a transparent medium with a radially dependent index of refraction, $n(r)$, similar to a crystal ball. Consequently, this equation is named accordingly. Under the assumptions made, the effective refractive index squared is obtained to the next to the leading order as follows:
\begin{equation}
n^{2}(r)=n^2_{\rm weak}(r)\approx 1+\frac{2m}{r}-\left(\frac{r_0}{r}\right)^\kappa,
\label{n2weak}
\end{equation}
where the second term represents gravitational lensing, while the third term represents the plasma de-lensing effect.

\begin{figure}
\vspace{15pt}
\center
\includegraphics[scale=0.9]{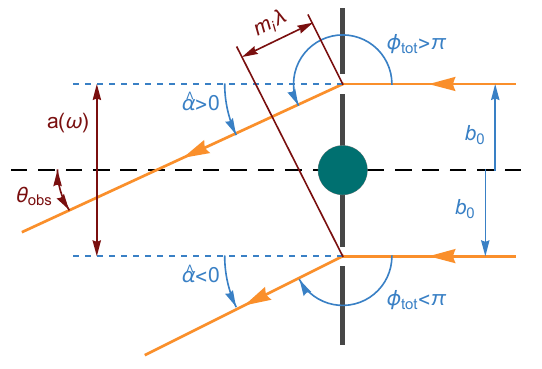}
\caption{A schematic representation of the pulsar system, illustrating its equivalence to Young's slits. Variables drawn with blue inc describe light ray propagation, while those in red inc describe two-slit interference.}  
\label{2slit}
\end{figure}

Next, we note an important point that the two nearly non-deflected rays, traveling on two opposite sides of the pulsar as in Fig. \ref{2slit}, can interfere, even at considerable distances. Such an interference process has been proposed \citep{M24} as an explanation for the multiple spectral bands (the so-called ``zebra pattern") observed in the dynamic spectra of the Crab pulsar high-frequency interpulse \citep{HE07, HE+16b, HE16c}. The system under consideration is therefore analogous to Young's two slits, as illustrated in Fig. \ref{2slit}. The separation between the ``slits" (referred to as ``hot spots" in \citealp{M24}) is twice the impact parameter distance at which the deflection angle vanishes, $a(\omega)=2b_{0}$.

Strictly speaking, the interference phenomenon does not necessitate the two rays to undergo minimal deflection. Large deflection angles are permissible for a generic Young's slit system. Nevertheless, observations of the Crab pulsar impose certain constraints on the possible deflection angle. We note that the deflection angle determines the difference in optical paths, $\Delta l$, and consequently the fringe order, $m_i$, as illustrated in Fig. \ref{2slit}. Specifically, constructive interference occurs when $a(\omega)\sin|\hat\alpha| = \Delta l = m_i\lambda$ (see discussion in the next section). For concreteness, let us assume that $a \sim 10^7$~cm (i.e., ten stellar radii) and $\lambda \sim 1$~cm (corresponding to a frequency of 30 GHz). Firstly, \citet{M24} shows that $m_{i}\gtrsim10^{2}$ is favored by observations, resulting in $|\hat\alpha|\gtrsim m_{i}\lambda/a\sim 10^{-5}$. Secondly, the duration of the HFIP pulses, $\tau$, is typically a few milliseconds. For interference to occur, the difference in light travel time along the rays should be much smaller than the pulse duration, $\Delta l \ll\tau c$. This yields the second constraint on the possible fringe order $m_{i}\sim\Delta l/\lambda\ll\tau c/\lambda\sim10^{4}$ and the deflection angle $|\hat\alpha|\sim\Delta l/a\ll 10^{-3}$. Consequently, the interfering rays must undergo a tiny, though non-zero, deflection.

By considering Eq. \eqref{cbe}, we observe that ray deflection vanishes identically when $n=1$. Indeed, direct integration with $n=1$ yields $\phi_{\rm tot}= \pi$ for any value of $b$. This is approximately the case for Eqs. \eqref{dfdr} and \eqref{ftot}, as well, provided $r_{m}\gg 2m,\, r_{0}$. We remind that the distance of the closed approach, $r_{m}$, is implicitly given by Eq. \eqref{rm}.  Furthermore, since the integral in Eq. \eqref{ftot} takes the major contribution near $r\sim r_{m}$ and noting that in the ``far-away approximation," 
\begin{equation}
r_{m}\simeq b_{0}, 
\label{rmb0}
\end{equation}
we can determine $b_{0}$ from
\begin{equation}
n_{\rm weak}(b_{0})\simeq 1.
\label{n=1}
\end{equation}
We remind that $b_{0}$ is the impact parameter of a special ray for which the defocussing effect of the plasma and the focussing effect due to gravity cancel each other. Upon solving Eq. \eqref{n=1}, we obtain
\begin{equation}
b_{0}^{\kappa-1}\simeq\frac{r_0^\kappa}{2m}=\frac{r_0^\kappa}{r_{\rm g}}.
\label{b0}
\end{equation}

Consequently, the slit (or hotspot) separation is 
\begin{equation}
a(\omega)=2b_{0}\simeq 2\left(\frac{r_0^\kappa}{r_{g}}\right)^{\frac{1}{\kappa-1}}\propto\omega^{-\frac{2}{\kappa-1}}.
\label{a-th}
\end{equation}
Hereafter, we omit the frequency subscript ``$\infty$'' as redundant.

\section{Interference}
\label{s-int}

The interference pattern generated by a monochromatic wave passing through two slits consists of bright fringes at locations where the phase difference along two paths is a multiple of $2\pi$, resulting in a path length difference that is a multiple of the wavelength, $\lambda$. The same ``striped" pattern is observed in the frequency domain by an observer located at a specific position, provided that the source possesses a broad-band spectrum.

For each frequency, the condition for a bright $i$-th fringe (interference maximum) is 
\begin{equation}
a(\omega_i)\sin\theta_{\rm obs}=m_{i}\lambda_{i}=[m_1\pm(i-1)](2\pi c/\omega_i),
\label{fringes}
\end{equation}
where $1\le i\le N$ and observations of the Crab pulsar's HFIP indicate the presence of approximately $N\sim30$ fringes between approximately 5~GHz and 30~GHz. Here, $m_{1}$ represents the fringe order (i.e., the number of the fringe) corresponding to the lowest frequency $\omega_{1}$. Observationally, the fringes exhibit a proportional separation with the ``6\%-rule," i.e., $\Delta\omega=(\omega_{i+1}-\omega_{i})\simeq0.057\,\omega_{i}$. Consequently, we define a parameter $\delta=0.057$ that characterizes the spectral band separation as follows:
\begin{equation}
\omega_{i+1}/\omega_{i}=(1+\delta).
\label{delta}
\end{equation}
Finally, $\theta_{\rm obs}$ denotes the location of an observer relative to the source-slits direction, as illustrated in Fig. \ref{2slit}. It is clear that $\theta_{\rm obs}=|\hat\alpha|$, with $\hat\alpha$ being the ray deflection angle. 

Combining Eqs. \eqref{fringes} and \eqref{delta}, we obtain the relation first derived in Ref. \citep{M24}:
\begin{equation}
a(\omega)=a(\omega_{1})\,\frac{\omega_1}{\omega}\left(1\pm\frac{\log_{1+\delta}(\omega/\omega_1)}{m_1}\right).
\label{a-exact}
\end{equation}
It is reasonable to assume that the total number of fringes produced in space significantly exceeds the quantity observed, $N_{\rm tot} \gg N$. Consequently, the order of any observed fringe, $m_{i}$, should also be substantially larger than $N$, as statistically $m_{i} \sim N_{\rm tot}/2$. This implies that there is a greater likelihood of observing higher-order fringes from the ``bulk" of the pattern compared to those near its edge, $m_{i} \lesssim N$. Therefore, assuming $m_{1} \delta \gg 1$, the above equation can be expressed in the following form:
\begin{equation}
a(\omega)=a(\omega_{1})\left(\frac{\omega}{\omega_1}\right)^{-\beta},\quad \beta=1\mp\frac{1}{m_1\delta}\approx1.
\label{a-obs}
\end{equation}

Direct comparison of Eqs. \eqref{a-th} and \eqref{a-obs} yields
\begin{align}
&\kappa=1+\frac{2}{1\mp\frac{1}{m_1\delta}}\approx 3, 
\label{kappa}\\
&a(\omega_{1})\approx2\left(\frac{r_{0}^{3}(\omega_{1})}{r_{g}}\right)^{1/2}
=2\left(\frac{r_{*}^{3}}{r_{g}}\right)^{1/2}\frac{\omega_{p*}}{\omega_{1}}
\label{aw1}
\end{align}
The plasma density profile derived from observations is, therefore,
\begin{equation}
n_{e}(r)\propto r^{-3}.
\label{ne-obs}
\end{equation}
This result is in remarkable agreement with theoretical expectations for a dipolar magnetic field.

Indeed, pulsar theory and simulations predict that the minimum density needed to be present in the magnetosphere for it to co-rotate with the pulsar as a rigid body up to the light cylinder is the so-called Goldreigh-Julian density,
\begin{equation}
n_{\rm GJ}(r)=\frac{\Omega B}{2ce}\simeq(0.07\textrm{ cm}^{-3})\frac{B}{P},
\end{equation}
where $B$ is the magnetic field strength in gauss and $P=2\pi/\Omega$ is the spin period in seconds. The actual density of plasma in the magnetosphere can be larger than $n_{\rm GJ}$ by a factor of multiplicity ${\cal M}>1$. For the Crab pulsar ($P\simeq0.0335$~s) with a dipolar field $B\simeq(4\times10^{12}\textrm{ G})(r/r_*)^{-3}$, 
\begin{equation}
n_{e}(r)\simeq (8.5 \times10^{12}\textrm{ cm}^{-3})\ {\cal M} \left(\frac{r}{r_*}\right)^{-3}
\propto r^{-3}.
\end{equation}
\begin{figure}[t]
\vspace{15pt}
\center
\includegraphics[scale=0.9]{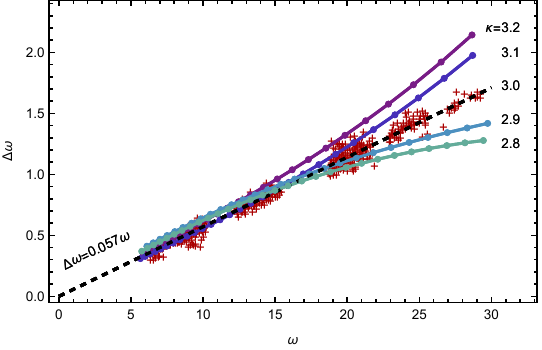}
\caption{The frequency fringe pattern obtained in the double-slit setup with the slit separation, $a(\omega)=2b_{0}$, obtained from Eq. \eqref{ftot} for $\phi_{\rm tot}=\pi$. Four different values of the power-law index $\kappa=2.8,\ 2.9,\ 3.1,\ 3.2$ are presented. The black dashed line indicates the anticipated dependence for the analytically obtained value of $\kappa=3$, see Eq. \eqref{kappa}. Other numerical parameters are adjusted so that there approximately $N\simeq30$ fringes in the 5--30 frequency domain (arb. units). The approximate data points from \citep{HE+16b} are also shown.
}
\label{zebra}
\end{figure}

For our analytical results obtained above, we employed the ``far-away approximation,'' which implies 
\begin{align}
&\frac{b_{0}}{r_{g}}=\left(\frac{r_{*}}{r_{g}}\right)^{3/2}\frac{\omega_{p*}}{\omega}\gg1,\\
&\frac{b_{0}}{r_{0}}=\left(\frac{r_{*}}{r_{g}}\right)^{1/2}\left(\frac{\omega_{p*}}{\omega}\right)^{1/3}\gg1.
\end{align}
For the Crab pulsar, the pulsar radius over the gravitational radius is $r_{*}/r_{g}\sim2.4$, and the plasma frequency at the star surface is 
\begin{equation}
\nu_{p*}=\omega_{p*}/2\pi\simeq 26{\cal M}^{1/2}\textrm{ GHz}. 
\end{equation}
Consequently, both constraints are reasonably well satisfied, with the exception of a small multiplicity, ${\cal M}\sim1$, when $\omega_{p*}\sim\omega$ near the upper end of the observational frequency range.

Now, we validate our findings directly from Eq. \eqref{ftot} without resorting to approximations. For this purpose, we numerically solve equation $\phi_{\rm tot}=\pi$ for $b_{0}$ across various values of $\kappa$. Notably, the obtained numerical solution, $b_{0}(r_{0})$, appears to be remarkably well approximated by the simplified analytical expression given by Eq. \eqref{b0}. Subsequently, employing $a=2b_{0}$, we compute the interference pattern and determine the positions of the fringes. The parameters of the model (e.g., the value of $m_{1}$) are properly adjusted to ensure that the number of fringes within the observational frequency range is about 30, as observed for the Crab pulsar. Finally, we present the $\Delta\omega$ vs. $\omega$ dependence in Fig. \ref{zebra}. In this figure, the results for $\kappa=2.8,\ 2.9,\ 3.1,\ 3.2$ are presented. Approximate data points from \citep{HE+16b} are also depicted. The straight black dashed line indicates the anticipated dependence for the analytically obtained value of $\kappa=3$ [given by Eq. \eqref{kappa}], which also coincides with the best fit to the data. It is evident that the numerical points converge to the observed linear trend $\Delta\omega=0.057\omega$ as $\kappa\to3$. Clearly, $\kappa\approx3$ provides the most accurate approximation to the observed data. In contrast, other values of $\kappa$ tend to deviate from the observed ``6\%-rule''.

\section{Discussion and predictions}
\label{s-disc}

(1) - Firstly, we observe that interference is capable of producing high-contrast fringe patterns, characterized by high visibility. The quantity called ``visibility'' serves as a measure of the contrast of a pattern and is defined as 
\begin{equation}
V = \frac{I_{\rm max} - I_{\rm min}}{I_{\rm max} + I_{\rm min}},
\end{equation}
where $I_{\rm max}$ and $I_{\rm min}$ represent the maximum and minimum intensities in the pattern. In our system, the path length difference between the two interfering rays, $m_{i}\lambda$, does not exceed tens of meters or less. This difference is orders of magnitude smaller than the system size. Consequently, the light attenuation (if any) in both rays should be nearly identical, leading to comparable wave amplitudes. Consequently, one can anticipate a nearly complete cancellation in the dark fringes, resulting in $I_{\rm min} \ll I_{\rm max}$. Hence, we obtain $V \simeq 1$, which aligns with observed data. 

It is instructive to emphasize the difference between interference and diffraction fringe patterns. In the latter, the modulation amplitude of the wave field diminishes with the fringe order, $m_{i}$, due to averaging over a large number of Fresnel zones, when $m_{i}\gg1$.

(2) - Another noteworthy item is the observed fluctuating dispersion measure excess, $\delta(\textrm{DM})\simeq0.02\textrm{ pc cm}^{-3}$ (at $\nu=10$~GHz), of the high-frequency interpulse compared to the main pulse. The dispersion measure is defined as the integral of the electron-positron density over a path length $l$, expressed as $\textrm{DM}\equiv\int n_{e}dl$. Essentially, it represents the column density of plasma along the line of sight. In our interference model, the excess DM is attributed to the propagation of the rays through the inner pulsar magnetosphere. The light rays traverse a dense region of the magnetosphere with $n_{e}\propto r^{-3}$. Consequently, the DM integral is predominantly influenced by the contribution near the distance of the closest approach $r_{m}\sim b_{0}$, where the density reaches its maximum. Hence, substituting the observed DM excess into equation
\begin{equation}
\delta(\textrm{DM})\simeq(0.91\textrm{ pc cm}^{-3}){\cal M}\left(\frac{b_{0}}{r_{*}}\right)^{-2},
\end{equation}
yields
\begin{equation}
b_{0}\simeq 6.7 {\cal M}^{1/2}r_{*}.
\end{equation}
This estimate is derived from the dispersion measure observations. An independent estimate is obtained from the interference model, as expressed by Eqs. \eqref{a-obs}, \eqref{aw1} [or Eq. \eqref{b0} with $\kappa\approx3$] with $\nu=10\textrm{ GHz}$.
\begin{equation}
b_{0}\approx\left(\frac{r_{*}}{r_{g}}\right)^{1/2}\frac{\nu_{p*}}{\nu}\simeq 4.2 {\cal M}^{1/2}r_{*}.
\end{equation}
The remarkable agreement between these two entirely independent estimates is truly noteworthy.

(3) - Next, our model assumes that the light rays propagate through the inner magnetosphere, hence $b_{0}\le r_{lc}$. The light cylinder is defined as the region where the linear speed approaches the speed of light, $\Omega\, r_{lc} \sim c$. Therefore, it serves as the outer boundary of the unperturbed dipolar corotating magnetosphere. The light cylinder radius is 
\begin{equation}
r_{lc}=\frac{c}{\Omega}\simeq(4.8\times10^{9}\textrm{ cm})P.
\end{equation}
For the Crab pulsar, $r_{lc}\simeq1.6\times10^{8}\textrm{ cm}\simeq 160 r_{*}$. 
Additionally, the ``hotspot distance,'' $b_{0}$, cannot be smaller than the neutron star radius, $b_{0}\ge r_{*}$, otherwise the light ray is absorbed by the surface. Altogether, these constraints take the form
\begin{equation}
1\lesssim \left(\frac{r_{*}}{r_{g}}\right)^{1/2}\frac{\omega_{p*}}{\omega}\lesssim160.
\end{equation}
These inequalities establish the frequency range within which the spectral ``zebra" pattern is anticipated to be observed.
\begin{equation}
(0.26\textrm{ GHz}){\cal M}^{1/2}\lesssim \nu \lesssim (42\textrm{ GHz}){\cal M}^{1/2},
\label{band}
\end{equation}
where we used that $(r_{*}/r_{g})^{1/2}\sim1.6$. It is important to note that the plasma multiplicity, ${\cal M}$, can exhibit variability with distance, rather than being a constant. Currently, there are no comprehensive models available that accurately describe the population and maintenance of plasma within the inner magnetospheres of pulsars. 

The lowest frequency at which the HFIP is observed is approximately 4 GHz, where it coexists with the low-frequency interpulse. The HFIP is not observed below this frequency but is observed at higher frequencies up to approximately 28 GHz, which is the maximum frequency used for the HFIP observations. We can see that the observed range, $4\textrm{ GHz}\le \nu \le28\textrm{ GHz}$, is entirely consistent with the constraint presented by Eq. \eqref{band}, thereby setting the constraint on plasma multiplicity:
\begin{equation}
0.44\lesssim{\cal M}\lesssim240.
\label{M}
\end{equation}
although, theoretically, multiplicity should not be lower than unity. 

(4) - We {\em propose to conduct observations} of the Crab pulsar HFIP and analyze its dynamic spectra at frequencies above 30~GHz, particularly in the millimeter and sub-millimeter ranges (e.g., with ALMA, SMA, other facilities). This is because the ``zebra" pattern is expected to undergo a change at a critical (maximal) frequency, $\nu_{c}$, at which the distance of the closest approach [and also $b_{0}$ by way of Eq. \eqref{rmb0}] is equal the neutron star's size $b_{0}\simeq r_{m}=r_{*}$, namely:
\begin{equation}
\nu_{c} \simeq (42\textrm{ GHz}){\cal M}^{1/2}.
\label{nuc}
\end{equation}
In general, the current model predicts that the critical frequency can be as low as $\nu_{c} \simeq 42\textrm{ GHz}$ (for ${\cal M}\simeq1$), although a larger value, up to $\nu_{c} \simeq 650\textrm{ GHz}$ (for ${\cal M}\simeq240$), cannot be ruled out.

At lower frequencies, $\nu < \nu_c$, the interference fringes are distinct and bright, with visibility approaching unity. Conversely, at higher frequencies, $\nu > \nu_c$, we have $b_{0}<r_{*}$, resulting in light rays encountering the surface of the neutron star and being absorbed. Consequently, only low-brightness refraction fringes can be observed, but their visibility is significantly diminished, $V \ll 1$. Given that the diffractive screen will be the bare neutron star of a constant size, the diffraction pattern will become equally and significantly more tightly separated, with the constant $\delta\nu$, as is estimated elsewhere \citep{M24}. The detection of such diffraction fringes may be quite challenging. 

Notably, the observation of the evolution of the spectral pattern with frequency and the identification of $\nu_{c}$ enable the deduction of the multiplicity, or equivalently, the magnetospheric plasma density at the star's surface for the first time. 

Furthermore, the ray path is influenced by both plasma dispersion and gravitational field. The interference is highly sensitive to variations in the ray path. Near and slightly below the critical frequency, light rays propagate close to the neutron star's surface, probing the strong field regime. We anticipate that the proportional spacing and the 6\%-rule may be altered by strong gravity. Consequently, distinguishing between the weak field estimate and the exact solution of the Crystal Ball equation, and comparing it with observational data can provide an independent test of general relativity.

(5) - Furthermore, we discuss a highly speculative hypothesis. \citet{M24} suggested that the high-frequency components, HFC1 and HFC2, observed at similar frequencies as HFIP but at different pulsar rotation phases, could be reflections off the magnetosphere of the same source that produces HFIP. In turn, \citet{HE+16b} assert that the high-frequency components are observed, albeit with limited confidence, down to 1.4 GHz. This frequency is well below the lowest frequency of 4~GHz, at which the HFIP becomes detectable. Within the framework of our reflection hypothesis, whereas light source is operational below the HFIP frequencies, the interpulse becomes visible only at such frequencies that allow for a straight, non-deflected propagation through the inner magnetosphere, that is when $b_{0}\lesssim r_{lc}$. Therefore,  4~GHz is the lower cut-off in Eq. \eqref{band}, which implies ${\cal M}\sim240$ and $\nu_{c}\sim650\textrm{ GHz}$. The primary limitation of this hypothesis lies in the assumption of an undisturbed dipolar field structure extending up to the light cylinder. In reality, this assumption is likely inaccurate, and the outer edge of the unperturbed dipole can be located at significantly smaller radii, $r < r_{lc}$. Consequently, both ${\cal M}$ and $\nu_{c}$ will be proportionally reduced.

(6) - Finally, we argue that our interference model imposes stringent constraints on the source's size, encompassing both lateral (spatial) and temporal coherence. Let us assume that the source possesses a transverse (perpendicular to the ray) size of $D$. Additionally, let us assume that each individual emission event, including a giant pulse (presumably hundreds or thousands of them comprise each microsecond-duration HFIP pulse), has a bandwidth $\delta\omega$. To ensure temporal coherence, i.e., the presence of interference, the condition $\delta\omega\, \tau_{c} \lesssim 1$ must be satisfied, where $\tau_{c}$ is the correlation time. For a typical sub-microsecond time scale, an estimate of $\delta\omega \lesssim 10^{7} \textrm{ s}^{-1}$ is obtained. The corresponding longitudinal coherence length $l\sim c\tau_{c}\sim 10^{3}$~cm.

To ensure spatial coherence, the light travel time disparity between the two interfering rays must result in a phase difference that is significantly less than $\pi$. Otherwise, fringes generated by the interference of light emitted at distances $D$ apart will overlap, resulting in a blurred pattern. Therefore, for a crisp interference pattern to be observed, the angular fringe size should be larger than the angular source size, $\lambda/a\gtrsim D/r_{lc}$. Assuming fiducial values of $\lambda \sim 3$~cm and $a \sim 10 r_* \sim 10^7$~cm, one estimates $D \lesssim 100$~cm. Both longitudinal and transverse scales are comparable to the inertial length (also called the skin scale, $d_{e}\equiv c/\omega_{p}$) near the light cylinder. Interestingly, such a source size is consistent with the radio wave emission mechanism via reconnection and plasmoid interaction \citet{P+19, PK22}.

\section{Conclusions}
\label{s-concl}

This paper investigates enigmatic radio observations of the high-frequency interpulse, which exhibit dynamic spectra characterized by a substantial number of proportionally spaced emission bands, colloquially referred to as the ``zebra" pattern. We elaborate upon the model proposed by \citet{M24}, wherein these bands are interpreted as interference fringes in the frequency domain, incorporating the effects of gravity. The model assumes that the source of HFIP is situated behind the pulsar, enabling light rays to propagate through the magnetosphere and interfere. In this study, we calculated the light ray geometry from first principles and directly deduced the properties of fringes. This model facilitated an accurate scan through the pulsar magnetosphere, enabling the precise determination of plasma parameters, such as plasma density, as a function of the magnetospheric radius. Consequently, we effectively performed a ``tomography" of the Crab pulsar magnetosphere. The deduced plasma density profile exhibits a power-law dependence, $n_e \propto r^{-3}$, which aligns remarkably with the predictions of the Goldreich-Julian theory for a dipolar magnetosphere. 

We observe that the combined effect of gravitational focusing and plasma defocusing of electromagnetic waves enables the creation of a high-contrast fringe pattern, characterized by the high visibility of order unity. This is in stark contrast to the dull and low-visibility diffractive pattern. Furthermore, our model appears to be the sole one that naturally accounts for all observed properties of the Crab pulsar HFIP radio observations, as outlined in the Introduction in seven bullet points. Our model also puts constraints on the source of the pulsar radio emission. 

Finally, we propose to conduct observations at frequencies exceeding 30~GHz. We anticipate that at a specific frequency, referred to as the critical frequency, the dynamic spectrum will undergo a modification. We predict a transition from the bright and high-contrast interference pattern to a significantly dimmer, very low-contrast, and closely spaced diffraction fringes. The detection of the latter may pose a substantial challenge. The critical frequency is predicted to be within a relatively narrow range, $42\text{ GHz}\lesssim \nu_{c}\lesssim 650\text{ GHz}$. The detection of this critical frequency would enable the complete identification of the density profile by determining the power-law normalization, specifically $n_{e}$ at the star's surface. Additionally, we note that the interference pattern itself is influenced by the space-time geometry, thereby providing an opportunity to explore gravity in the strong field regime, particularly near the star's surface.

\section*{Acknowledgements}
The author thanks Roger Blandford, Ramesh Narayan, Avi Loeb for useful and insightful comments. He is also grateful to the organizers (Uzdensky, Philippov, Spitkovsky, Willingale) and participants of the KITP 2025 program on ``Relativistic plasma physics'' for stimulating discussions. The author is also very grateful to Dmitri Uzdensky for his very careful reading of the manuscript, and for providing constructive, insightful, and valuable comments. 
The author acknowledges support by the National Science Foundation (NSF) under grant PHY-2409249. 
This research was supported in part by grant NSF PHY-2309135 to the Kavli Institute for Theoretical Physics (KITP). 

\section*{Declaration of interests}

The authors report no conflict of interest.

\appendix

Consider a transparent medium with a radially dependent index of refraction, denoted as $n(r)$. Our objective is to determine the geometric path traversed by light within this medium. Given the spherical symmetry of such a medium, reminiscent of a spherical glass or crystal, we refer to it as the ``Crystal Ball equation." Conservation of angular momentum implies that light motion is restricted to a plane, $\theta = const.$, which can be arbitrarily chosen as $\theta = {\pi}/{2}$. Consequently, we seek the ray path expressed in terms of $r$ versus $\phi$, or vice versa.

In geometric optics, the light path is determined by the shortest travel time, so the variation $\delta t=0$, where
\begin{equation}
t=\int \frac{dl}{v}=\frac{1}{c}\int n(r)\, dl=\frac{1}{c}\int n(r)\sqrt{1+r^{2}(\phi')^{2}}\, dr,
\end{equation}
with $\phi'= d\phi/dr$ and $dl^{2}=dr^{2}+r^{2} d\phi^{2}$. The common constant $c$ can be omitted. It is observed that the expression within the integral represents the Lagrangian of the system,
\begin{equation}
L(r,\phi)=n(r)\sqrt{1+r^{2}(\phi')^{2}}.
\end{equation}

The Euler-Lagrange equation 
\begin{equation}
\frac{d}{dr}\left(\frac{\partial L}{\partial\phi'}\right)-\frac{\partial L}{\partial \phi}=0,
\end{equation}
can be readily integrated once because the second term vanishes identically. This yields
\begin{equation}
n(r)\, \frac{r^{2}\phi'}{\sqrt{1+r^{2}(\phi')^{2}}}=b,
\end{equation}
where $b$ is a constant of integration. Rearranging the terms, we obtain the Crystal Ball equation:
\begin{equation}
\frac{d\phi}{dr}=\frac{b}{r\sqrt{r^{2} n^{2}(r)-b^{2}}}.
\end{equation}
In this context, the physical meaning of the constant $b$ is that it represents the impact parameter.

It is important to note that Eq. \eqref{dfdr}, derived within the framework of general relativity, incorporates an additional factor in the denominator, specifically the  ``lapse function," $\sqrt{1-2m/r}$. This factor is absent in classical optics. In the weak field approximation, the contribution of this factor becomes negligible, leading to its omission in the leading order, as in Eq. \eqref{cbe}.


\bibliographystyle{aasjournalv7}
\bibliography{refs}

\end{document}